\documentclass{iopart}

\usepackage{amsthm}
\usepackage{amssymb}
\usepackage{caption}
\usepackage{cite}
\usepackage{algorithm}
\usepackage{algpseudocode}
\usepackage{color,graphicx,subfigure}
\usepackage{multirow}
\usepackage{booktabs}

\usepackage{iopams}
\expandafter\let\csname equation*\endcsname\relax
\expandafter\let\csname endequation*\endcsname\relax
\usepackage{amsmath}
\usepackage[english]{babel}
\usepackage{hyperref}

\newcommand{\zmark}[1]{{#1}}

\begin{document}
	\title[Joint bi-modal image reconstruction]{Joint bi-modal image reconstruction of DOT and XCT with an extended Mumford-Shah functional}
	
	\author{Di He$^1$\footnote{Part of the work was conducted while Di He was a PH.D student of Peking University.} , Ming Jiang$^{2,3}$, Alfred K. Louis$^{4}$, Peter Maass$^{5}$, Thomas Page$^6$\footnote{Part of the work was conducted while Thomas Page was a PH.D student of University Breman and visiting student of Peking University.}  }
	
	\address{$^1$School of Applied Science, Beijing Information Science \& Technology University, Beijing 100192, China.}
	\address{$^2$ LMAM, School of Mathematical Sciences, Peking University, Beijing 100871, China.}
	\address{$^3$ Beijing International Center for Mathematical Research, and Cooperative Medianet Innovation Center, Peking University, Beijing 100871, China.}
	\address{$^4$ Institute of Applied Mathematics, Saarland University, 66041 Saarbr\"{u}cken, Germany.}
	\address{$^5$ Center for Industrial Mathematics, University of Bremen, 28359 Bremen, Germany.}
	\address{$^6$ Automated Driving Lab China, BMW China Services LTD.}
	
	\ead{dihe@bistu.edu.cn, ming-jiang@pku.edu.cn, louis@num.uni-sb.de, pmaass@math.uni-bremen.de, Thomas.TP.Page@bmw.com.}

	\begin{abstract}
		Feature similarity measures are indispensable for joint image reconstruction in  multi-modality medical imaging, which enable joint multi-modal image reconstruction (JmmIR) by communication of feature information from one modality to another, and vice versa. In this work, we establish an image similarity measure in terms of image edges from Tversky's theory of feature similarity in psychology. For joint bi-modal image reconstruction (JbmIR), it is found that this image similarity measure is an extended Mumford-Shah functional with {\it a priori} edge information proposed previously from the perspective of regularization approach.  This image similarity measure consists of Hausdorff measures of the common and different parts of image edges from both modalities. By construction, it posits that two images are more similar if they have more common edges and fewer unique/distinctive features, and will not force the nonexistent structures to be reconstructed when applied to JbmIR. With the $\Gamma$-approximation of the JbmIR functional, an \zmark{alternating} minimization method is proposed for the JbmIR of diffuse optical tomography and x-ray computed tomography. The performance of the proposed method is evaluated by three numerical phantoms. It is found that the proposed method improves the reconstructed image quality by more than $10\%$ compared to single modality image reconstruction (SmIR) in terms of the structural similarity index measure (SSIM)
	\end{abstract}
	
	\vspace{2pc}
	\noindent{\it Keywords}: Single-modal image reconstruction (SmIR), joint multi-modal image reconstruction (JmmIR), joint bi-modal image reconstruction (JbmIR),  image similarity measures, extended Mumford-Shah functional, diffuse optical tomography (DOT), x-ray computed tomography (XCT).

\maketitle

\section{Introduction}

Biomedical imaging aims at visualizing structural or functional information necessary for medical research and clinical diagnosis. Each imaging modality can provide images of one particular physical, physiological, or biological distribution. Multi-modality medical imaging technique is to combine multiple imaging modalities into one hybrid imaging system, such as PET/MRI \cite{Judenhofer2008, Catana2017}, PET/XCT  \cite{Cherry2009, Bockisch2009, Delbeke2009, Even2009, Kaufmann2009}, SPECT/XCT \cite{Cherry2009, Bockisch2009, Delbeke2009, Even2009, Kaufmann2009}, XCT/MRI \cite{Wang2015}, DOT/MRI \cite{Panagiotou_2009} and DOT/XCT \cite{Yuan_2010, Boas2014, Deng2015, Baikejiang2017}, BLT/DOT/XCT \cite{Yang2015}, and an  omni-tomography system of multiple modalities \cite{Wang2012}.\footnote{The abbreviations are as follows in the order of appearance: positron emission tomography (PET), magnetic resonance imaging (MRI), x-ray computed tomography (XCT), single-photon emission computed tomography (SPECT), diffuse optical tomography (DOT), bioluminescence tomography (BLT).} Hybrid multi-modality imaging systems can visualize anatomical and functional structures simultaneously, improve the overall imaging performance, especially avoiding the tempo-spatial artifacts because of scanning with different devices at different time and positions \cite{Townsend2008}. It offers significant diagnostic advantages that cannot be achieved by a single modality \cite{Steiner2007}. Moreover, if hybrid multi-modality imaging systems are perfectly calibrated, image registration is not necessary. Multi-modality medical imaging within one hybrid imaging system is called \emph{hardware fusion} in \cite{Townsend2008}.

For single-modal image reconstruction (SmIR), theories as well as algorithms have already been well developed \cite{Natterer_2002, Scherzer_2010, Censor_2008}. Image reconstructions for hybrid multi-modality imaging systems can be conducted in several ways. The first approach is that all the image reconstructions are conducted separately by \zmark{multiple} SmIRs without sharing information across imaging modalities, which is a simple application of multiple SmIRs for each modality. The second approach is that all the image reconstructions are conducted sequentially by using reconstructed images to guide the next image reconstruction. The sequential reconstructions are usually started from high resolution imaging modalities such as XCT or MRI. The structural information of reconstructed images is used to guide the next reconstructions by applying structural similarity. This approach is based on the observation that images of the same object possess similar structural information, especially at key structures or strong image edges, though with different contrasts from different modalities, because they are images of the same anatomical structure. This approach is called \emph{model fusion} in \cite{Haber2013} or \emph{software fusion} in \cite{Townsend2008}. The third approach is that all the image reconstructions are jointly conducted by sharing reciprocally structural information from one modality to another, and vice versa. This is called \emph{joint inversion} in  \cite{Haber2013} and  joint multi-modal image reconstruction (JmmIR) in this paper. For the joint bi-modal image reconstruction such as DOT/XCT in this paper, it is called joint bi-modal image reconstruction (JbmIR).

JmmIR can be achieved by using iteratively model fusion with an appropriate image similarity measure of structural information in certain feature space for images of the underlying modalities \cite{Haber2013}. Given such an image similarity measure, the image structural feature from one modality can guide the  reconstruction and help to improve the image quality of another modality, and vice versa, in one joint reconstruction process, mostly in an alternatively iterative manner. Model fusion only applies the structural information from one reconstructed image unidirectionally to improve the next reconstruction, but not vice versa. Unlike model fusion, the structural information of images in a JmmIR process is not from an image already reconstructed, but is built up progressively during the joint reconstruction process and applied reciprocally among images of underlying modalities. It is expected that JmmIR can be jointly performed with enhanced image quality but less measured data via the reciprocal communication of structural feature information. Nevertheless, the success of JmmIR relies on both the representation of image structural features of different modalities and the similarity measure between the features.

Image edge is the fundamental representation of structural information of image \cite{marr_1982}. In this work, we derive an image similarity measure of image edges from Tversky's theory of feature similarity in psychology and apply it to the JbmIR for DOT and XCT. For JbmIR, it is found that the derived image similarity measure is relevant to the extended Mumford-Shah functional with {\it a priori} edge information proposed previously for model fusion from the perspective of regularization approach in \cite{Page_2015}. This image similarity measure consists of Hausdorff measures of the common and different parts of image edges from DOT and XCT, and can be approximated with edge indicator functions by the $\Gamma$-convergence theory. With the $\Gamma$-approximation of the JbmIR functional, an \zmark{alternating} minimization method is proposed for the JbmIR of DOT and XCT. The performance of the proposed method is evaluated with three numerical phantoms. It is found that the proposed method improves the image quality by more than $10\%$ in terms of the structural similarity index measure (SSIM) for both XCT and DOT.

The structure of the paper is as follows. We first review the previous work on model fusion and JmmIR in \S~\ref{sec:related:work}. Then in \S~\ref{sec:similarity:measure} we derive an image similarity measure of image edges from Tversky's theory of feature similarity in psychology and propose a method for JbmIR based on image edge in \S~\ref{sec:JbmIR_edges}.  In \S~\ref{sec:approx:JbmIR} we propose a heuristic $\Gamma$-approximation for the proposed JbmIR method. In \S~\ref{sec:XCT:DOT:problem} we apply the JbmIR method to the JbmIR of XCT and DOT with an alternating minimization algorithm for the implementation of the method. In \S~\ref{sec:numerical}, we perform numerical experiments with three phantoms with single modal image reconstruction with the Mumford-Shah (SmIR-MS) regularization with our JbmIR algorithm (JbmIR-MS). We discuss the results, relevant issues and future work in \S~\ref{sec:discussion}. We conclude this paper in \S~\ref{sec:conclusion}.

\section{Related work}
\label{sec:related:work}

Since JmmIR can be achieved by using iteratively model fusion, we review the previous methods for model fusion and discuss the challenges for JmmIR in this section. The following classification of methods is based on the major characteristics of methods, and is used to address the challenges for JmmIR, but is not intended to be a strict classification. There are certainly overlaps among the classifications, e.g., derivatives or its discretization, finite differences, are used explicitly or implicitly in almost all the methods.

\subsection{Segmentation-based model fusion}
\label{subsec:seg:based}
For model fusion with MRI or CT, image edges from MRI or CT are used to replace the line process in the Gibbs distribution of \cite{Geman_1984}. This approach is first used for the reconstruction of PET or SPECT \cite{Leahy_1991, Fessler1992, Gindi_1993}. If the edge information from MRI is imperfect and if the edge process is improperly weighted, using such information ``blindly'' can even lead to artifacts \cite{Fessler1992, Zhang1994, Bowsher_1996}. In \cite{Bowsher_1996}, a simultaneous segmentation and reconstruction method for emission computed tomography (ECT) image by incorporating high-resolution anatomical information such as CT or MRI is proposed and evaluated with SPECT and MRI. Higher prior probabilities are assigned to ECT segmentations in which each ECT region stays within a single anatomical region in this incorporation \cite{Bowsher_1996}.  In \cite{Sastry_1997} , a different method for the incorporation of anatomical information into PET image reconstruction is proposed: segmentations of MRI images are used to assign tissue composition to PET image pixels, while PET images are modeled as the sum of activities of each tissue type, weighted by the assigned tissue composition.

If segmentation from a high resolution imaging modality is available, {\it a priori} structural information is used for DOT, as the weight for an anisotropic regularization to preserve the edges in the DOT image under reconstruction in \cite{Douiri2007}. The idea of \cite{Douiri2007} is later applied for PET in \cite{Chan2009} and for SPECT in \cite{Kazantsev2012, Dewaraja2010}. For the model fusion of DOT with XCT, a similarity measure with a Laplacian-type smoothing operator within each region is proposed to average the DOT image within a region, respectively, while allowing discontinuity between different regions. \cite{Brooksby2005, Brooksby2006, Yalavarthy2007, Baikejiang2017}. In \cite{Fang2010, Boas2014, Deng2015}, an empirical compositional relationship between the x-ray intensities for adipose and fibroglandular tissue is further proposed with the segmentation information from XCT, and used to guide the reconstruction of DOT in \cite{Fang2010, Boas2014, Deng2015}. \footnote{To be precise, it is the x-ray tomosynthesis, a tomographic technique of limited views,  that is used in \cite{Fang2010, Boas2014, Deng2015}, instead of the conventional XCT with full-scanned completed data. }

This approach requires both \emph{accurate anatomical image segmentation and registration} \cite{Fessler1992, Gindi_1993, Ouyang_1994, Sastry_1997}.

\subsection{Segmentation-free model fusion}
\label{subsec:seg:free}
To resolve the difficulty in accurate anatomical image segmentation, information theoretic regularization using the mutual information or joint entropy to measure image similarity of image intensities are studied in \cite{Nuyts2007,Tang2009, Panagiotou_2009, Somayajula_2011}. MRI is used as the information theoretic anatomical prior for the reconstruction of PET in \cite{Somayajula_2011} and DOT in \cite{Panagiotou_2009}. It is found that the joint entropy is more robust than the mutual information to differences between the image under reconstruction and the anatomical image provided, especially when the anatomical image has structures that are not present in the image under reconstruction \cite{Nuyts2007, Tang2009, Panagiotou_2009, Somayajula_2011}.

However, because both information theoretic metrics are based on global distributions of the image intensities, mutual information or joint entropy between images is not directly affected by local spatial structure of the images. ``{\it A random spatial reordering of corresponding pairs of voxels in the two images would produce identical measures}'' \cite{Somayajula_2011}. Therefore, It is necessary to incorporate features that capture the local variation in image intensities. In \cite{Somayajula_2011}, scale-space features, such as a Gaussian-blurred image and its Laplacian, are added into the similarity measure to help the joint entropy reconstruction. This method is related to the following derivative-based approach.

Another disadvantage of information theoretic regularization is that it is prone to different image contrast ranges from different modalities and may lead to undesired reconstruction results \cite{Panagiotou_2009, Somayajula_2011,Weizman2016}.

\subsection{Derivative-based model joint reconstruction}
 \label{subsec:derivative:based}
As mentioned in the introduction, unlike model fusion, the structural information in a JmmIR process is not from an image already reconstructed, but is updated progressively during the joint reconstruction process. Most of the model fusion methods in \S~\ref{subsec:seg:based}~and \zmark{\S~\ref{subsec:seg:free}} have not been adopted to JmmIR, because of challenges in progressively computing image segmentations and quantifying similarity of image structural features of different contrast ranges from different modalities.

Another image representation, the gradient of an image serves as a convenient image edge detector and is independent of image contrast ranges \cite{marr_hldreth_1980, canny_1986, perona_malik_1990}. The gradient and higher order derivatives such as Laplacian of an image provide alternative structural representations to resolve the aforementioned issues. They are easy to be updated progressively. With derivatives as the image structural representation, the problem is how to define the similarity measure of image structures.

This approach was first applied in geophysical imaging from multiple physical process \cite{Haber1997}.  In \cite{Haber1997}, the $L^2$-norm of second order derivatives of image differences is used as the similarity measure for the joint inversion of gravity and seismic tomography data. A generalized cross-gradient procedure is developed for joint multiple parameter inversion in \cite{gallardo_2007, gallardo_2011}, where cross-products of imaging gradients are utilized to measure the parallelism of gradients, a.k.a., the structural similarity of physical parameters.

The total variation is one of the most widely used regularizations in image reconstruction problems \cite{rudin_et_al_1992}. For JmmIR, the joint total variation by summing up the total variations of images is used to solve the joint inversion problem in \cite{Haber2013}. However, such a joint total variation encourages the joint sparsity but not the similarity of image gradients. In \cite{Rigie2015}, the total nuclear variation (TNV) is proposed for encouraging common edge locations and alignment of their gradient vectors, and is applied to the multi-spectral CT reconstruction. In \cite{Knoll_2017}, the second order total generalized variation (TGV) with nuclear-norm is applied for the joint image reconstruction of MRI and PET. It is found that the TGV nuclear norm is robust regarding unwanted transfer of individual features to other channels \cite{Knoll_2017}.

In \cite{Ehrhardt_2015}, for the JbmIR of PET ($u_1$) and MRI ($u_2$), another image similarity measure for image gradients is proposed to measure the parallelism of image gradients as follows, called parallel level sets
(PLS),
\begin{equation}
\label{eq:PLS}
\text{PLS}_\beta (u_1, u_2)
=
\int_{\Omega}
\varphi
\left[
\psi\left(\|\nabla u_1\|_\beta \cdot\|\nabla u_2\|_\beta \right)
-
\psi\left(|\langle\nabla u_1,\nabla u_2\rangle|_{\beta^2} \right)
\right]
\end{equation}
where $\Omega\subset \mathbb{R}^{N}$ is a domain that contains the object to be imaged, for $N = 2$ or $3$,  $\|z\|_\beta = (\|z\|^2+\beta^2)^{\frac{1}{2}}$ for some $\beta > 0$ for the differentiability of $\text{PLS}_\beta$, $\langle\cdot,\cdot\rangle$ is the Euclidean inner product, $\phi$ and $\psi$ are strictly increasing functions. Please note that this image similarity measure is symmetric in $u_1$ and $u_2$.  Please note that asymmetric PLS can be formulated along the same framework \cite{Ehrhardt2015thesis}.

However, the above derivative-based methods depend on the magnitude of gradients. For JmmIR, images tend to have different ranges of contrasts and different heights and signs of edges detected with derivatives \cite{rasch_2017}. Normalized PLS can be formulated to resolve this issue within the framework in \cite{Ehrhardt2015thesis}. Another approach to resolve this issue is using the infimal convolution of generalized Bregman distances to formulate weighted similarity measure for image gradients without the edge orientation constraint \cite{rasch_2017}, where asymmetric weightings are also possible. This approach has also the advantage for joint reconstruction to avoid the artificial transfer of non-shared structures between the images \cite{rasch_2017}.

\subsection{Summary}
\label{subsec:jmmir:summary}
All work reviewed above are valuable contributions because they demonstrate the performance and advantages of JbmIR over SmIR in a number of multi-modal imaging applications. All the methods in this subsection use image gradients or derivatives as the image structural representation. \zmark{Similarity} measures are proposed to measure the parallelism or correlation of image gradients or derivatives, respectively. In spite of their demonstrated success, none of them is working directly on image edge. For image reconstruction, feature reconstruction \cite{louis_2015} provides the possibility for directly addressing structural information on features like image edges, but has not yet \zmark{been} used for joint multi-modal image reconstruction. It is our motivation to develop an approach directly with image edge together with an appropriate similarity measure, which is expect to be more robust to data noise than the derivative-based methods.

As discussed in \cite{Haber2013}, the challenges in JmmIR are as follows. The first fundamental challenge is because of the difference of contrast ranges or different scales of different imaging modalities \cite{rasch_2017}, or different physical meaning (and units) \cite{Haber2013}. Because structural similarity is the only clue for JmmIR, the second one is an appropriate representation of image structural information, which should support to capture the local variation in image intensities, be independent of image contrasts, and be updated progressively during the joint reconstruction process. Consequently the third one is an appropriate similarity measure for the representation of image structural information. There could be features in one modality but nonexistent in another. Hence, an appropriate similarity measure should encourage the reconstruction of features only if there is sufficient evidence from measured data support but should not force the nonexistent features to be reconstructed in another.

In this work, image edge is used as the structural representation of images, which is the line-process in the Gibbs distribution of \cite{Geman_1984}, and edge set \zmark{in} the Mumford-Shah functional \cite{Mumford_1989}. This representation meets the criteria of the second challenge aforementioned. A similarity measure for image edge sets \zmark{is} derived from the feature contrast model of similarity in psychology by Tversky \cite{Tversky_1977}, which is appropriate in the sense of the third challenge aforementioned.

\section{Similarity measures for image edges}
\label{sec:similarity:measure}

Many similarity measures explain similarity (or dissimilarity) as a distance $d(\cdot, \cdot)$ in  a metric space $\mathcal{X}$ of perceptual stimulus of objects, with $d$ satisfying the following metric axioms \cite{Tversky_1977},
\begin{enumerate}
\item Minimality: $d(K_1,K_2) \geq d(K_1, K_1) = 0$.
\item Symmetry: $d(K_1,K_2) = d (K_2,K_1)$.
\item Triangle inequality: $d(K_1,K_2) + d(K_2,K_3) \geq d(K_1,K_3)$,
\end{enumerate}
where $K_1$, $K_2$ and $K_3 \in \mathcal{X}$. This is called the geometric model of similarity \cite{Blough2001, Rorissa2007}. However, all three metric axioms are unnecessarily stringent and can be inconsistent with a number of psychological experiments \cite{Tversky_1977, Mumford1991, Rorissa2007, Skopal2011, Biasotti2016}. It is realized that ``Similarity isn't a metric anyway'' in \cite{Mumford1991} after knowing the work of \cite{Tversky_1977}.

In his prominent paper \cite{Tversky_1977}, Tversky proposed his famous feature contrast model for the similarity of binary features. Let $\Delta = \{a,b,c, \cdots\}$ be the domain of objects under study. Assume that each object in $\Delta$ is represented by a set of binary features. Unlike the geometric model of similarity which represent stimulus as points in a metric space, an object $a$ is characterized by the set $A$ of features that the object $a$ possesses. Features are binary in the sense that a given feature of an object either is or is not in its set of features $A$. An example of the set of features is the image edge set of an image. Tversky proposed another set of axioms about similarity measures of binary features, which \zmark{include} the axioms of matching, independence, solvability, invariance, and proved mathematically that feature similarity measures must be of the following form \cite{Tversky_1977}\footnote{Interested readers are referred the paper \cite{Tversky_1977} for details. Although the axioms in \cite{Tversky_1977} are translated in some papers, we found that the original presentation in \cite{Tversky_1977} is still the best. Please note that some translations are incomplete due to the complicated formulations of axioms, especially for the axioms of solvability and invariance. It is for the same reason that we give up the translation of Tversky's axioms in this paper.}
\begin{equation}\label{eq:feature:contrat}
S(a, b)= p(A\cap B) - \gamma_1 p(A\setminus B) - \gamma_2 p(B\setminus A),
\end{equation}
where $A$ and $B$ denote the sets of binary features associated with the objects $a$ and $b$, respectively,  $\gamma_1$ and $\gamma_2$ are nonnegative constants. $p$ is an additive function such that $p(A\cap B) =  p (A) + p(B)$ whenever $A \cap B = \emptyset$. This representation is called the feature contrast model and has been validated with multiple psychological experimental data sets \cite{Tversky_1977, Tversky_1982} and applied in image processing applications \cite{Santini1997, Santini_1999, Rorissa2007}. Please note that the convention here is that the more similar $A$ to $B$, the bigger the similarity measure $S(A,B)$.  This formula \eqref{eq:feature:contrat} implies that a reasonable feature similarity measure is a linear combination of the measures of their common and their distinct features.

Similarity measures thus obtained increase with addition of common features and/or deletion of distinctive features (i.e., features that belong to one object but not to the other) \cite{Tversky_1977}. They posit that two stimuli are more similar if they have more common features and fewer unique/distinctive features \cite{Rorissa2007}. Hence, similarity measures will not force the nonexistent features to be reconstructed when applied to JmmIR as discussed in \S~\ref{subsec:jmmir:summary}. Actually, this point is postulated as the matching axiom in \cite{Tversky_1977}. Psychologically, this explains why subjects tend to pay more attention to common features than distinctive features in their similarity judgments \cite{Tversky_1977}.

The similarity in \zmark{\eqref{eq:feature:contrat}} is asymmetric when \zmark{$\gamma_1 \neq \gamma_2$}. It should address that asymmetric similarity measures are in general necessary for JmmIR. Because of different image contrasts among modalities, similarity measures should be asymmetric at least from a heuristic perspective. Some of the approaches in \S~\ref{sec:related:work} \zmark{provide} asymmetric similarity measures with demonstrated advantages. At this point, It is necessary to address one distinct approach for the model fusion of PET from MRI in \cite{Bowsher2004}, which has not \zmark{been} reviewed in  \S~\ref{sec:related:work}. It is a segmentation-free method with a smoothing Markov prior with weights determined by the MRI image intensities at nearby voxels. It can be used in an asymmetrical way by using only partial derivatives in certain directions \cite{Bowsher2004, Schramm2018}. It is reported in \cite{Schramm2018} that the PLS is slightly inferior compared to the asymmetrical Bowsher method.

For 2D images, a natural choice for the measure $p$ of image edge sets is the 1-dimensional Hausdorff measure $\mathcal{H}$, or the length of image edges. Therefore, we obtain,
\begin{equation}\label{eq:feature:contrat:haus}
S(u_1,u_2)=
\mathcal{H}(K_1\cap K_2) - \gamma_1 \mathcal{H}(K_1\setminus K_2) - \gamma_2 \mathcal{H}(K_2\setminus K_1),
\end{equation}
where $K_1$ and $K_2$ denote the image edge sets of images $u_1$ and $u_2$, respectively. This similarity measure of image edges is closely related to the Mumford-Shah regularization functional \cite{Mumford_1989}. Higher-dimensional Hausdorff measures can also be used in \eqref{eq:feature:contrat:haus} for higher-dimensional images \cite{Jiang_2014}.

\section{JbmIR based on image edges}
\label{sec:JbmIR_edges}

In this section, we apply the similarity measure for image edges of the previous section to JbmIR and establish a general approach for JbmIR based on image edges. Without loss of generality, we restrict to the bi-modal case and 2D case in this paper to avoid notational complexity, although the mathematical formulation can be extended to general multi-modal case.

For multi-modality imaging systems, the image reconstruction problem is to estimate images $u_i$ from measurement data $g_i$, for $i = 1, 2$,
 \begin{equation}\label{eq:image:rec}
 A_i (u_i)=g_i,
 \end{equation}
 where $A_i$ is the forward operator of underlying imaging modality. The Mumford-Shah functional was first proposed as a variational approach for image denoising and segmentation from the Markov random field theory \cite{Geman_1984}. It provides an approach for incorporating edges for the regularization of image reconstruction problems.   Recently the Mumford-Shah functional has been successfully applied as a regularization method for image reconstruction problems \eqref{eq:image:rec}\cite{Rondi_2001,Ramlau_2007,Rondi_2008,Klann_2011,Jiang_2014}. The SmIR for \eqref{eq:image:rec} with the Mumford-Shah regularization is as follows,
\begin{equation}
\label{eq:image:rec:MS:1}
\text{MS}(u_i, K_i)
=\int_{\Theta_i}|A_i(u_i)-g_i|^2+\alpha_i\int_{\Omega\setminus K}|\nabla u_i|^2+\beta_i\mathcal{H}(K_i),
\end{equation}
where $\Theta_i$ is the domain of the measurement, $K_i$ is the edge set of the image $u_i$,  $\alpha_i$ and $\beta_i$ are positive regularization parameters, for $i = 1, 2$.

The JbmIR based on image edges is to use the similarity measure in \eqref{eq:feature:contrat:haus} to formulate the following joint reconstruction functional
\begin{align}
E(u_1,u_2,K_1,K_2)
= \text{MS}(u_1, K_1) + \text{MS}(u_2, K_2) - \tau S(u_1,u_2),
\end{align}
where $\tau > 0$ is a regularization parameter. Please note that the minus sign ``$-$'' is to ensure the similarity of edge sets of images $u_1$ and $u_2$ during the joint reconstruction process because of the convention that the more similar edges, the bigger their similarity. Let
\begin{align}
\Phi_i (u_i, K_i) = & \int_{\Theta_i}|A_i(u_i)-g_i|^2 + \alpha_i \int_{\Omega\setminus K_i}|\nabla u_i|^2.
\end{align}
Then we find,
\begin{multline}
E(u_1,u_2,K_1,K_2) \\
=  \Phi_1(u_1, K_1) + \Phi_2(u_2, K_2)
 + \beta_1 \mathcal{H}(K_1) + \beta_2 \mathcal{H}(K_2)  \\
 - \tau \mathcal{H}(K_1\cap K_2)
 + \tau \gamma_1 \mathcal{H}(K_1\setminus K_2)
 + \tau \gamma_2 \mathcal{H}(K_2\setminus K_1).
\end{multline}
In the above equation, the terms involving only the edges $K_1$ and $K_2$ \zmark{sum} up to
\begin{align}
\label{eq:edge:reg:1}
   & \beta_1 \mathcal{H}(K_1) + \beta_2 \mathcal{H}(K_2)
   - \tau \mathcal{H}(K_1\cap K_2)
   + \tau \gamma_1 \mathcal{H}(K_1\setminus K_2)
    + \tau \gamma_2 \mathcal{H}(K_2\setminus K_1)\\
=&
    \beta_1 \mathcal{H}(K_1 \cap K_2) + \beta_1 \mathcal{H}(K_1 \setminus K_2)
    + \beta_2 \mathcal{H}(K_2 \cap K_1) + \beta_2 \mathcal{H}(K_2 \setminus K_1) \\
  & - \tau \mathcal{H}(K_1\cap K_2)
    + \tau \gamma_1 \mathcal{H}(K_1\setminus K_2)
    + \tau \gamma_2 \mathcal{H}(K_2\setminus K_1)\\
\label{eq:edge:reg:4}
=&
    (\beta_1 + \beta_2 - \tau) \mathcal{H}(K_1 \cap K_2) + (\beta_1 + \tau \gamma_1) \mathcal{H}(K_1 \setminus K_2)
    +  (\beta_2 + \tau \gamma_2) \mathcal{H}(K_2 \setminus K_1).
\end{align}
By re-parametrization with the reuse of the same notations for the regularization parameters $\beta_1$, $\beta_2$, $\gamma_1$, and $\gamma_2$ to avoid notational complexity, this sum can be written as the following,
\footnote{Let $\beta_1'$, $\beta_2'$, $\gamma_1'$, and $\gamma_2'$ be positive parameters such that
\begin{align}
\beta_1 + \beta_2  - \tau & = \beta_1' + \beta_2',\\
\beta_1 + \tau \gamma_1 & = \beta_1'  \gamma_1',\\
\beta_2 + \tau \gamma_2 & = \beta_2'  \gamma_2.'
\end{align}
Here we assume that $\beta_1 + \beta_2  > \tau$.  This is a reasonable assumption because all the edge regularizations in \eqref{eq:edge:reg:4} reduce to the Mumford-Shah regularization on image edge with the regularization parameter $\beta_1 + \beta_2  - \tau$, in the ideal case when the image edge sets of $u_1$ and $u_2$ perfectly match, i.e., $K_1 = K_2$. The equation \eqref{eq:similarity:Psi} holds with $\beta_1$, $\beta_2$, $\gamma_1$, and $\gamma_2$ being replaced by $\beta_1'$, $\beta_2'$, $\gamma_1'$, and $\gamma_2'$, respectively.}
\begin{multline}
\label{eq:similarity:Psi}
\Psi(K_1,K_2) \\
=
    \beta_1\left[
        \mathcal{H}(K_1\cap K_2)+\gamma_1\mathcal{H}(K_1\setminus K_2)
    \right]
    +
    \beta_2\left[
      \mathcal{H}(K_1\cap K_2)+\gamma_2\mathcal{H}(K_2\setminus K_1)
    \right].
\end{multline}
In summary, the  joint reconstruction functional can be written as
\begin{multline}\label{eq:JbmIR:K}
E(u_1,u_2,K_1,K_2) \\
=
\sum_{i=1, j \neq i}^2
\int_{\Theta_i}|A_i(u_i)-g_i|^2 + \alpha_i \int_{\Omega\setminus K_i}|\nabla u_i|^2
+
\beta_i\left[
        \mathcal{H}(K_i\cap K_j)+\gamma_i\mathcal{H}(K_i\setminus K_j)
    \right],
\end{multline}
where $\alpha_i$, $\beta_i$ and $\gamma_i$ ($i = 1, 2$) are positive regularization parameters.

\section{Variational approximation for JbmIR}
\label{sec:approx:JbmIR}

In \cite{Page_2015}, an extended Mumford-Shah functional with {\it a priori} edge information was established for image reconstruction from the perspective of regularization theory. For an image reconstruction problem $A(u) = g$ with similar notations as in \eqref{eq:image:rec} and \eqref{eq:image:rec:MS:1} , if partial information of image edge can be obtained, then the following reconstruction functional is proposed in \cite{Page_2015},
\begin{multline}\label{eq:image:rec:MS:K0}
\text{MS}_{K^0}(u,K)\\
=\int_{\Theta}|A(u)-g|^2 +\alpha\int_{\Omega\setminus K}|\nabla u|^2
+\beta\left[\mathcal{H}(K\setminus K^0)+\gamma\mathcal{H}(K\cap K^0)\right],
\end{multline}
where $K_0$ is the partial information of image edge. Theoretical results and numerical experiments were reported in \cite{Page_2015}. The advantages of the extended Mumford-Shah functional were demonstrated by numerical experiments with the model fusion of DOT from XCT.

By ignoring the accurate values of parameters, we have the joint reconstruction functional in \eqref{eq:JbmIR:K}
\begin{equation}\label{E:MS1:MS2}
E(u_1,u_2,K_1,K_2) = \text{MS}_{K_2}(u_1,K_1) + \text{MS}_{K_1}(u_2,K_2).
\end{equation}
However, because none of $K_1$ and $K_2$ is fixed, it is difficult to extend the theoretical results in \cite{Page_2015} on existence, regularization property, $\Gamma$-approximation to the joint reconstruction functional. JbmIR based on image edge as formulated in \eqref{eq:JbmIR:K} is to find the minimizer pair $\left(u_1,u_2,K_1,K_2\right)$ given the measurement $g_1$ and $g_2$ and appropriate choice of the regularization parameters. As its native form for SmIR as in \eqref{eq:image:rec:MS:1}, the joint reconstruction functional is not easy to minimize numerically because of the difficulty of tractable edge representation in implementation. $\Gamma$-approximation is an approach to find a variational approximation for minimizing the the joint reconstruction functional. We are to propose an approximation for the joint reconstruction functional and demonstrate its performance with numerical experiments in this work, based on previous work in \cite{Page_2015}.

In \cite{Page_2015}, the following $\Gamma$-approximation to $\text{MS}_{K^0}(u,K)$ in \eqref{eq:image:rec:MS:K0} is proposed,
\begin{multline}\label{similarity:Page15:Gamma:approx}
\text{MS}_{K^0}(u, v) \\
= \int_{\Theta_i}|A(u)-g|^2+\alpha\int_{\Omega}v^2|\nabla u|^2
+\beta
\left[
\int_{\Omega}\left(\varepsilon|\nabla v|^2
+\frac{(1-v)^2}{4\varepsilon}\right)(1+\gamma(v_0-v)^2)\right],
\end{multline}
where $0 \le  v \le 1$ and $0 \le v_0 \le 1$ are the edge indicator functions for the edge sets $K$ and $K_0$, respectively, i.e., $v(x) = 1$ if $x \notin K$ and $v(x) = 0$ if $x \in K$, with the same definition for $v_0$ and $K_0$. $\varepsilon$ controls the ``width'' of edges. The $\Gamma$-convergence of $\text{MS}_{K^0}(u, v)$ is only proved for $\gamma = 0$ in \cite{Page_2015}. Although the $\Gamma$-approximation in \eqref{similarity:Page15:Gamma:approx} is proposed from an heuristic perspective, its performance is demonstrated with numerical experiments in \cite{Page_2015}. Based on the relation in \eqref{E:MS1:MS2}, we propose the following $\Gamma$-approximation to the joint reconstruction functional $E$ in \eqref{eq:JbmIR:K} as follows,
\begin{multline}    \label{similarity:Psi:Gamma:approx}
F(u_1,v_1,u_2,v_2)\\
=
\sum_{i=1, j \neq i}^2
\int_{\Theta_i}|A_i(u_i)-g_i|^2+\alpha_i\int_{\Omega}v^2_i|\nabla u_i|^2
+\beta_i
    \left[
        \int_{\Omega}\left(\varepsilon_i|\nabla v_i|^2
    +\frac{(1-v_i)^2}{4\varepsilon_i}\right)(1+\gamma_i(v_j-v_i)^2)
    \right],
\end{multline}
where $0 \le  v_1 \le 1$ and $0 \le v_2 \le 1$ are the edge indicator functions for the edge sets $K_1$ and $K_2$, respectively. $\varepsilon_i$ controls the ``width'' of edges. As $\varepsilon_1 \to 0+$ and $\varepsilon_2 \to 0+$, the proposed $\Gamma$-approximation \eqref{similarity:Psi:Gamma:approx} is expected to $\Gamma$-converge to the functional in \eqref{eq:similarity:Psi}, and the edge indicator functions $v_1$ and $v_2$ are expected to converge to the characteristic functions of the edge sets $K_1$ and $K_2$, respectively, in the sense of $L^2$.  The current formulation of the $\Gamma$-approximation \eqref{similarity:Psi:Gamma:approx} is based on the heuristics in \cite[\S~4.5]{Page_2015}. We are unable to prove the expected convergence mathematically, although a proof for $\gamma_1 = \gamma_2 = 0$ might be possible by using the same method in \cite{Page_2015} where a lengthy proof for single modal image reconstruction with the extended Mumford-Shah function with {\it a priori} edge information is provided for $\gamma = 0$. In this work we concentrate in studying numerically how the proposed method can help \zmark{to} improve the quality of reconstructed images.

\section{JbmIR of XCT and DOT}
\label{sec:XCT:DOT:problem}

In this section we give a brief introduction of XCT and DOT, and apply the JbmIR reconstruction method in the previous section to the JbmIR of XCT and DOT.

\subsection{X-ray computed tomography}

In conventional X-ray tomography, the image contrast is due to the x-ray absorption when X-ray beams pass through an object. The interaction of X-ray and the object could be a complex process. Nevertheless, the beam path in straight lines provides a good approximation for X-ray tomography in many cases. The absorption process is described by Beer's law. Please refer to \cite{Natterer_2001, louis1989inverse} for more details.

Let $I(x)$ be the intensity of an X-ray and $f(x)$ the X-ray attenuation coefficient at position $x$. Then along a straight line $L$,
\begin{equation}\label{eq:Beer:law}
\frac{d I}{d l}=-f d l,
\end{equation}
where $d l$ is the differential element along $L$. Let $I_0$ be the initial intensity of the X-ray and $I_1$ the intensity after passing through the object. As we assume that the beam travels in the straight line $L$, from \eqref{eq:Beer:law} it follows
\begin{equation}
\ln \frac{I_1}{I_0} = -\int_{L} f dl.
\end{equation}
The measurement $I_1$ and the initial intensity $I_0$ provide line integrals of the X-ray attenuation coefficient $f$ along lines, if the measurement is conducted in multiple directions and positions. The operator that maps a function into the set of its line integrals is the Radon Transform, which is defined as the following in the case of 2D XCT,
\begin{equation}
R(f) (\theta, s) = \int_{L} f dl,
\end{equation}
for every line $L$ crossing the domain $\Omega \subset \mathbb{R}^2$ containing underlying objects, where $\theta =  (\cos \varphi, \sin \varphi)$ is the normal direction of the line $L = \{ (x, y) \in \mathbb{R}^2:\, x \cos \varphi + y \sin \varphi = s\}$ and $s \in  \mathbb{R}$. Let $\Theta_1 = S^1 \times \mathbb{R}$ (where $S^1$ is the unit circle), and $u_1 = f$, and $g_1 = \ln \frac{I_1}{I_0}$. Then the XCT is to estimate the image $u_1$ from its measurement data $g_1$,
\begin{equation}
R(u_1) = g_1.
\end{equation}

\subsection{Diffuse optical tomography}

The steady-state diffuse optical tomography (DOT) is a functional imaging modality which uses near infrared light to illuminate biological tissue and reconstruct the inside distribution of optical properties using the light intensity measured on the surface of the tissue. Light at near-infrared wavelength will be absorbed and scattered while passing through biological tissue. The optical properties have direct biological relevance since the absorption of light is due to the existence of oxy-/deoxy-haemoglobin. Please refer to   \cite{Arridge_1999,Arridge_2009} for more details.

The governing equation used for diffuse light $u$ in DOT is the diffuse approximation of the radiative transport equation (RTE),
\begin{equation}
\label{eq:DOT}
-\hbox{div}(D\nabla u)+\mu_a u=0, \qquad \text{in $\Omega$}
\end{equation}
where $D$ is the diffusion coefficient and $\mu_a$ is the absorption coefficient. The incoming light $q$ can be modeled by Robin boundary condition in the diffuse approximation,
\begin{equation}\label{eq:DOT:source}
u+2D \overrightarrow{n}\cdot\nabla u=q, \qquad \text{on $\partial\Omega$}
\end{equation}
where $\overrightarrow{n}$ is the outer normal on the boundary $\partial\Omega$ of $\Omega$. The measurement $g_2$ is the negative Neumann boundary values of the solution of equation \eqref{eq:DOT}
\begin{equation}\label{eq:DOT:measure}
g_2=-D \overrightarrow{n}\cdot\nabla u, \qquad \text{on $\partial\Omega$}.
\end{equation}
In practice, extra coupling coefficients due to refraction should be introduced into the equations \eqref{eq:DOT:source} and \eqref{eq:DOT:measure} \cite{Arridge_1999,Arridge_2009}.  Given the incoming light $q$, diffusion coefficient $D$ and absorption coefficient $\mu_a$, the forward operator $\mathcal{F}$ in DOT maps $(\mu_a, D)$ to the measurement data $g_2$ in \eqref{eq:DOT:measure}. The DOT is to estimate $(\mu_a, D)$ from the measurement data $g_2$,
\begin{equation}
\mathcal{F}(\mu_a,D)=g_2.
\end{equation}

In this work we consider a special case of DOT to demonstrate the performance of the proposed JbmIR method. We are interested in recovering the absorption coefficient $\mu_a$, and assume the diffusion coefficient $D$ to be known in the following. This is an interesting but non-linear and ill-posed inverse problem \cite{Arridge_2009}. For a known diffusion coefficient $D$,  we introduce the new forward operator $G$ as
\begin{equation}
G(\mu_a)=\mathcal{F}(\mu_a,D)
\end{equation}
To use the same notations in the previous sections, we write $u_2$ to represent the absorption image $\mu_a$, and $g_2$ for the measurement $G(\mu_a)$. Then the DOT with know diffuse coefficient is to estimate $u_2$ from the measurement data $g_2$ ,
\begin{equation}
G(u_2)=g_2,
\end{equation}
where the measurement is on the boundary $\Theta_2 = \partial\Omega$ of $\Omega$.

\subsection{JbmIR of XCT and DOT}

With the development of \zmark{hybrid} multi-modal imaging techniques, there are a number of \zmark{hybrid} optical tomography systems reported in the literature together with reconstruction methods by model fusion from XCT or JbmIR with XCT, such as DOT/XCT \cite{Yuan_2010, Boas2014, Deng2015, Baikejiang2017}, BLT/DOT/XCT \cite{Yang2015}. To reduce the radiation dose and accelerate the data acquisition, it is the x-ray tomosynthesis technique of limited views that is used in the latter work \cite{Baikejiang2017, Boas2014, Deng2015} rather than the conventional full-scan XCT.

By applying the general JbmIR formulation in \eqref{similarity:Psi:Gamma:approx}, we obtain the following joint reconstruction function for XCT and DOT,
\begin{multline}    \label{JbmIR:XCT:DOT}
F(u_1,v_1,u_2,v_2)\\
=
\int_{\Theta_1}|R(u_1)-g_1|^2+\alpha_i\int_{\Omega}v^2_1|\nabla u_1|^2
+\beta_1
    \left[
        \int_{\Omega}\left(\varepsilon_1|\nabla v_1|^2
    +\frac{(1-v_1)^2}{4\varepsilon_1}\right)(1+\gamma_1(v_2-v_1)^2)
    \right] \\
+
\int_{\Theta_2}|G(u_2)-g_2|^2+\alpha_2\int_{\Omega}v^2_2|\nabla u_2|^2
+\beta_2
    \left[
        \int_{\Omega}\left(\varepsilon_2|\nabla v_2|^2
    +\frac{(1-v_2)^2}{4\varepsilon_2}\right)(1+\gamma_2(v_1-v_2)^2)
    \right],
\end{multline}
where $0 \le  v_1 \le 1$ and $0 \le v_2 \le 1$ are the edge indicator functions for the edge sets of XCT image $u_1$ and DOT image $u_2$, respectively. Please note that there are no weighting parameters in front of the two fidelity terms, but this does not mean both fidelity terms contribute equally to the joint reconstruction. Because of the \zmark{alternating} minimization algorithm discussed in the following, they never meet each other and their contributions are independent of each other. The difference of their contribution are weighted by the regularization parameters $\beta_1$, $\beta_2$, $\gamma_1$ and $\gamma_2$. During the training of the regularization parameters, the two fidelity terms also play roles independent of each other. The difference of their contribution, even if there are different weights for both, will merged into the regularization parameters.

The JbmIR of XCT and DOT in this work is to reconstruct the images $u_1$ and $u_2$ as well as the edge indicators $v_1$ and $v_2$ simultaneously in an iterative process. This is achieved by minimizing the objective functional \eqref{JbmIR:XCT:DOT} with an \zmark{alternating} minimization algorithm in the order of $u_1$, $v_1$, $u_2$ and $v_2$. Each minimization for $u_1$, $v_1$, $u_2$ and $v_2$ is a gradient descending with the backtracking line search method by the Armijo-Goldstein stopping condition \cite{Armijo_1966, wiki:Backtracking}, with the previous reconstruction as initial value and the others being frozen at the latest reconstructed values. Hence, the proposed \zmark{alternating} minimization algorithm for the JbmIR of XCT and DOT is an instance of stochastic gradient descending with the aforementioned order, and is described in Algorithm~\ref{AL:whole}. Notice that we use $\arg\min$ in the algorithm description to indicate our minimizing purpose and in each step we use gradient descent method, therefore the results we get from the iteration are the approximations of \zmark{minimizers}.
\begin{algorithm}[htbp]
	\caption{\zmark{Alternating} minimization for JbmIR of XCT and DOT}
	\label{AL:whole}
	\begin{algorithmic}
		\For{$i=0,1,2,\ldots,n$}
		\begin{align}
            \label{eq:XCT:MS:u}
			u_1^{i+1}=&\arg\min_{u_1} F(u_1,v_1^i,u_2^i,v_2^i) \\
            \nonumber
			&\hbox{with $u_1^i$ as the initial value};\\
            \label{eq:XCT:MS:v}
			v_1^{i+1}=&\arg\min_{v_1} F(u_1^{i+1},v_1,u_2^i,v_2^i) \\
            \nonumber
			&\hbox{with $v_1^i$ as the initial value};\\
            \label{eq:DOT:MS:u}
			u_2^{i+1}=&\arg\min_{u_2}F(u_1^{i+1},v_1^{i+1},u_2,v_2^i)\\
            \nonumber
			&\hbox{with $u_2^i$ as the initial value};\\
            \label{eq:DOT:MS:v}
        	v_2^{i+1}=&\arg\min_{v_2}F(u_1^{i+1},v_1^{i+1},u_2^{i+1},v_2)\\
            \nonumber
            &\hbox{with $v_2^i$ as the initial value};
		\end{align}
		\EndFor
	\end{algorithmic}
\end{algorithm}

Although there are many terms in the joint reconstruction functional $F$ in \eqref{JbmIR:XCT:DOT}, there is not \zmark{so many} computing terms at each minimization step in \eqref{eq:XCT:MS:u}, \eqref{eq:XCT:MS:v} \eqref{eq:DOT:MS:u} and \eqref{eq:DOT:MS:v} because some terms can be treated as constants. For example, at the minimization step \eqref{eq:XCT:MS:u}, only the first two terms at the first line on the right-hand side of \eqref{JbmIR:XCT:DOT} are involved in this minimization step, those terms not involving $u_1$ can be treated as constants and ignored. Especially the time consuming computation of the forward process $G$ for the DOT is not needed. We have carefully analyzed the dependent terms at each minimization step to reduce the computing load. Implementation and experimental details are reported in the next section.

\section{Numerical experiments}
\label{sec:numerical}

In discrete case, the gradients are discretized with forward difference and the divergence with backward difference, so that the discretized gradient corresponds to the gradient of the discretized functional.

For XCT, the initial image value is set as zero and the initial edge value is set as one. For DOT, zero and one initials cannot lead to good reconstruction results. We use gradient \zmark{descent} method for only the fidelity term to get a \zmark{blurry} image as the initial image value, and use the Mumford-Shah functional segmenting this initial image to get the initial edge value. The SmIR-MS and JbmIR-MS both use this same initial value setting.

We also need to set up the stopping criterion of our algorithm. This is achieved by setting up the iteration numbers involved. After a number of numerical experiments with phantoms of similar structures as in this work, we find that after 80 whole iterations with 10 steps for each variables in every iteration the SSIM values, which we use to evaluate the quality of the reconstructed images, of both XCT and DOT stabilize at relatively higher values, and in our examples over 0.85 for XCT and over 0.69 for DOT.

We evaluate the quality of the reconstructed image $u_{rec}$ via the structural similarity index measure (SSIM)\cite{Wang_2004} using the corresponding true image $u_{true}$ as a reference image:
\begin{equation}
\hbox{SSIM}(u_{rec},u_{true})=\frac{(2\overline{u}_{true}\overline{u}_{rec}+C_1)(2\sigma_{u_{true}u_{rec}}+C_2)}{(\overline{u}_{true}^2+\overline{u}_{rec}^2+C_1)(\sigma_{u_{true}}^2+\sigma_{u_{rec}}^2+C_2)}
\end{equation}
where $\overline{u}_{true}$, $\overline{u}_{rec}$ are averages, $\sigma^2_{u_{true}}$, $\sigma^2_{u_{rec}}$ are variances and $\sigma_{u_{true}u_{rec}}$ is the covariance.

SSIM value is proposed to measure the similarity between two images considering three comparison measurements: luminance, contrast and structure. We use the default parameter settings that $C_1=(0.01L)^2$, $C_2=(0.03L)^2$, where $L$ is the dynamic range of the ground truth. The higher SSIM value is the more similar the two images are. SSIM=1 only happens for identical images.

\subsection{Experimental setting}
The computations are done using Matlab. For XCT, the Radon Transformation and the adjoint operator written by Lut Justen from the Software-Documentation of the Center for Industrial Mathematics, University of Bremen are implemented. For DOT, the TOAST++ package\cite{Schweiger_2014} from Martin Schweiger and Simon Arridge is implemented for the forward operator and the Jacobian matrix of the forward operator.

\begin{figure}[!htbp]
	\centering
	\includegraphics[width=8cm]{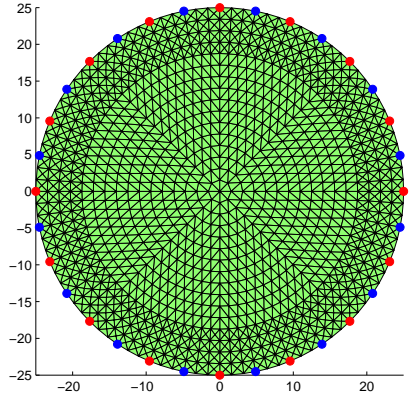}\\
	\caption{Finite element mesh for DOT problem with 16 source positions (red points) and 16 detector positions (blue points).}\label{fig:mesh}
\end{figure}

For XCT, the projection data \zmark{are} obtained from 30 views uniformly distributed in $[0,180^o]$.  Because our phantoms are of radius 25 mm and image sizes are of $100\times100$ pixels, the package we use for XCT automatically assigns each view 100 uniformly distributed parallel X-ray beams with a step-length of 0.5 mm. For DOT, the data \zmark{are} obtained with 16 source positions and 16 detector positions. TOAST++ uses finite element method to solve DOT problem. The mesh and the positions of source and detector are shown in Figure \ref{fig:mesh}.

\zmark{We add additional Gaussian noise $n$ to our noise free data $d$. Since the fidelity term $\int_{\Theta_i}|A_i(u_i)-g_i|^2$ shows that $A_i(u_i)$ should be a good approximation of the data $g_i$ in $L_2$ norm, we scale the noise $n$ to the data with regard to the $L_2$ norm. The relative noise level is defined as $\eta=\frac{\|n\|}{\|d\|}$, where $\|\cdot\|$  is the Euclidean norm. }In our numerical experiments 5\% relative Gaussian noise is added to the XCT data and 2\% relative Gaussian noise is added to the DOT data.

\subsection{Parameter choosing}
\label{subsec:parameter}

The choice of the regularization \zmark{parameters} is non-trivial and should be done task-dependently \cite{Schramm2018}.

In our proposed JbmIR-MS method, the ``width'' control parameters $\varepsilon_1$ and $\varepsilon_2$ are set as $1\times 10^{-4}$, which are the same as in \cite{Page_2015}. There are three regularization parameters $\alpha_i$, $\beta_i$ and $\gamma_i$ $(i=1,2)$ needed to be chosen properly for each modality. $\alpha_i$ is the regularization parameter of the smoothing penalty term. The larger it is, the smoother the reconstructed image will be when there is not any edge. $\beta_i$ weights how strongly the edge penalizes. The larger it is, the fewer edges will be reconstructed. $\gamma_i$ controls the effect of the reference edge information. The effect of the reference edge information increases as the value of $\gamma_i$ increases.

For JbmIR-MS, from the XCT with \emph{a priori} edge information numerical experiments done by Thomas we can get that when the order of magnitude of $\alpha_1$ is around $10^4$, the order of magnitude of $\beta_1$ is around $10^{-3}$ or $10^{-4}$ and $\gamma_1$ is no more than $10$ we could be able to get XCT results with clear structures. We first fix the XCT parameters as $\alpha_1=9\times10^3,~\beta_1=2\times10^{-3},~\gamma_1=10$ to choose DOT parameters. For joint DOT reconstruction we first roughly partition the parameter space and compute the SSIM value to find rough-optimal parameters. Then according to the reconstruction results and the parameter effect discussed above, we fine-tune the parameters around this rough-optimal choice. For example, if the reconstructed image is too smooth in non-edge area, we will decrease the value of $\alpha_2$. If there are too few edges, we will slightly decrease the value of $\beta_2$. If the influence of the reference edge information is too strong that we reconstruct wrong structures that should only appear in one modality and not in the other, we will slightly decrease the value of $\gamma_2$. Here $\alpha_2$ is chosen from $A_2=[0.01,0.1,\ldots,1\times 10^9]$, $\beta_2$ is chosen from $B_2=[1\times 10^{-9},1\times 10^{-8},\ldots,0.1,1]$, and $\gamma_2$ is chosen from $C_2=[1,2,\ldots,9,10]$. \zmark{Theses parameter ranges} are chosen from our experience with a number of phantoms of similar structures. For our numerical experiments, we find that $\alpha_2=1\times10^5$, $\beta_2=1\times10^{-5}$ and $\gamma_2=5$ produce a higher SSIM value of the reconstructed DOT image against its true image, among all possible combinations from $A\times B \times C$. Then we fine-tune this rough-optimal choice according to the discussion above and reach the parameter settings we use in our numerical experiments. After choosing the DOT parameters, we then fine-tune the joint XCT parameters around $\alpha_1=9\times10^{3},~\beta_1=2\times10^{-3},~\gamma_1=10$. The final parameter settings for every experiment are listed in the description of the numerical results.

We compare our method with SmIR using Mumford-Shah functional(SmIR-MS).

For SmIR-MS, the ``width'' control parameter $\varepsilon$ is also set as $1\times10^{-4}$. There are two regularization parameters $\alpha$ and $\beta$ for each modality. The reconstruction result is sensitive to both of them. The larger $\alpha$ is the smoother the reconstructed image will be when there is not any edge. The larger $\beta$ is the fewer edges will be reconstructed. We first perform a search of the parameter space in a rough scale with the SSIM as the criterion. For both single modality reconstructions $\alpha$ and $\beta$ are chosen from $A=[0.01,0.1,\ldots,1\times 10^9]$ and $B=[1\times 10^{-9},1\times 10^{-8},\ldots,0.1,1]$ to determine the proper order of magnitudes first. For DOT reconstruction we find that $\alpha=1\times 10^5$ and $\beta=1\times 10^{-7}$ produce a higher SSIM value of the reconstructed DOT image against its true image, among all possible combinations from $A\times B$. For XCT reconstruction the parameters are $\alpha=1\times 10^4$ and $\beta=1\times10^{-2}$. We then fine-tune this rough-optimal choice according to the discussion above and our final parameter settings for every experiment are listed in the description of the numerical results.

\subsection{Phantom setting}
\begin{figure}[!htbp]
	\centering
	\includegraphics[width = 10cm]{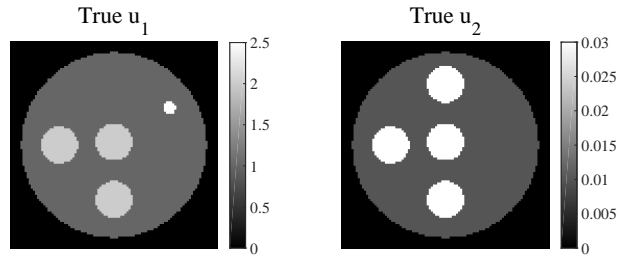}
	\caption{Phantom 1. Left: the true distribution of the X-ray attenuation coefficient $u_1$.  Right: the true distribution of the absorption coefficient $u_2$.}
	\label{fig:truedistribution}
\end{figure}

\begin{figure}[!htbp]
	\centering
	\includegraphics[width = 10cm]{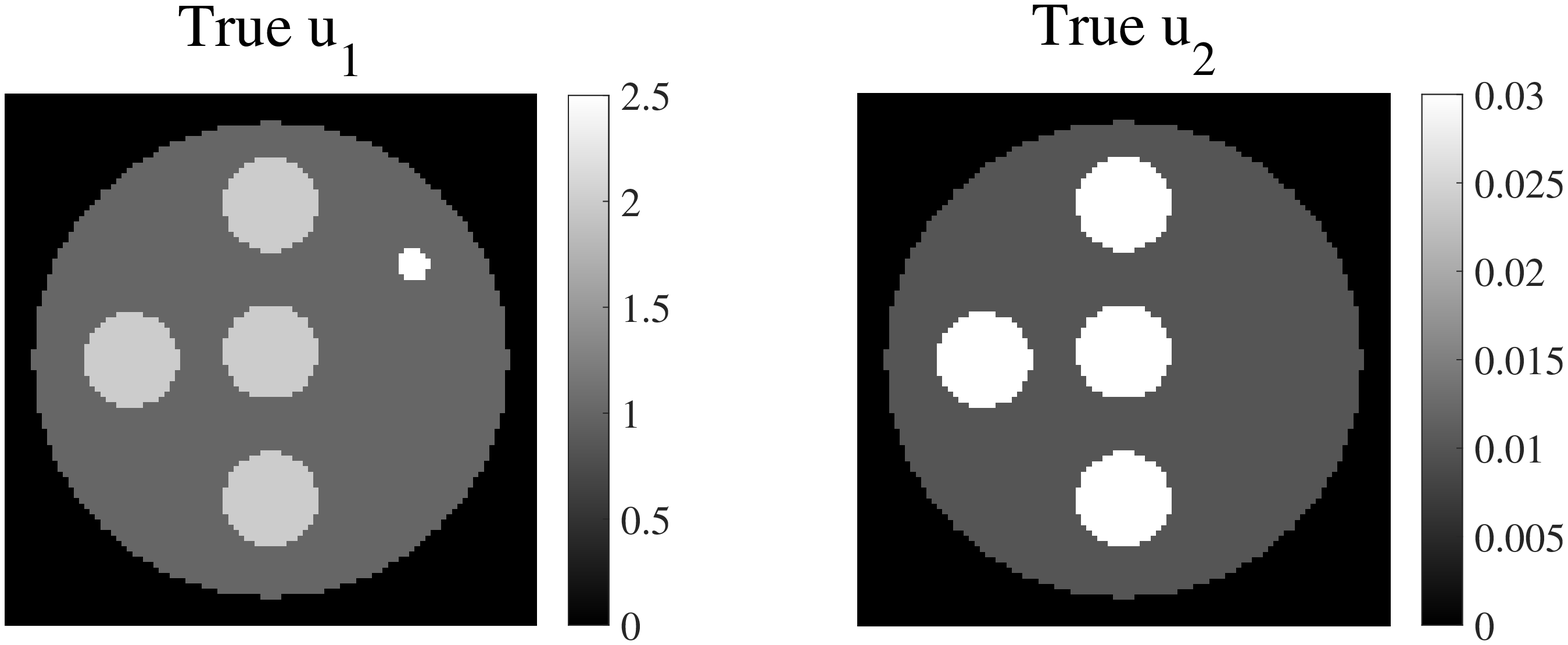}
	\caption{Phantom 2. Left: the true distribution of the X-ray attenuation coefficient $u_1$. Right: the true distribution of the absorption coefficient $u_2$.}
	\label{fig:truedistribution_same}
\end{figure}

\begin{figure}[!htbp]
	\centering
	\includegraphics[width = 10cm]{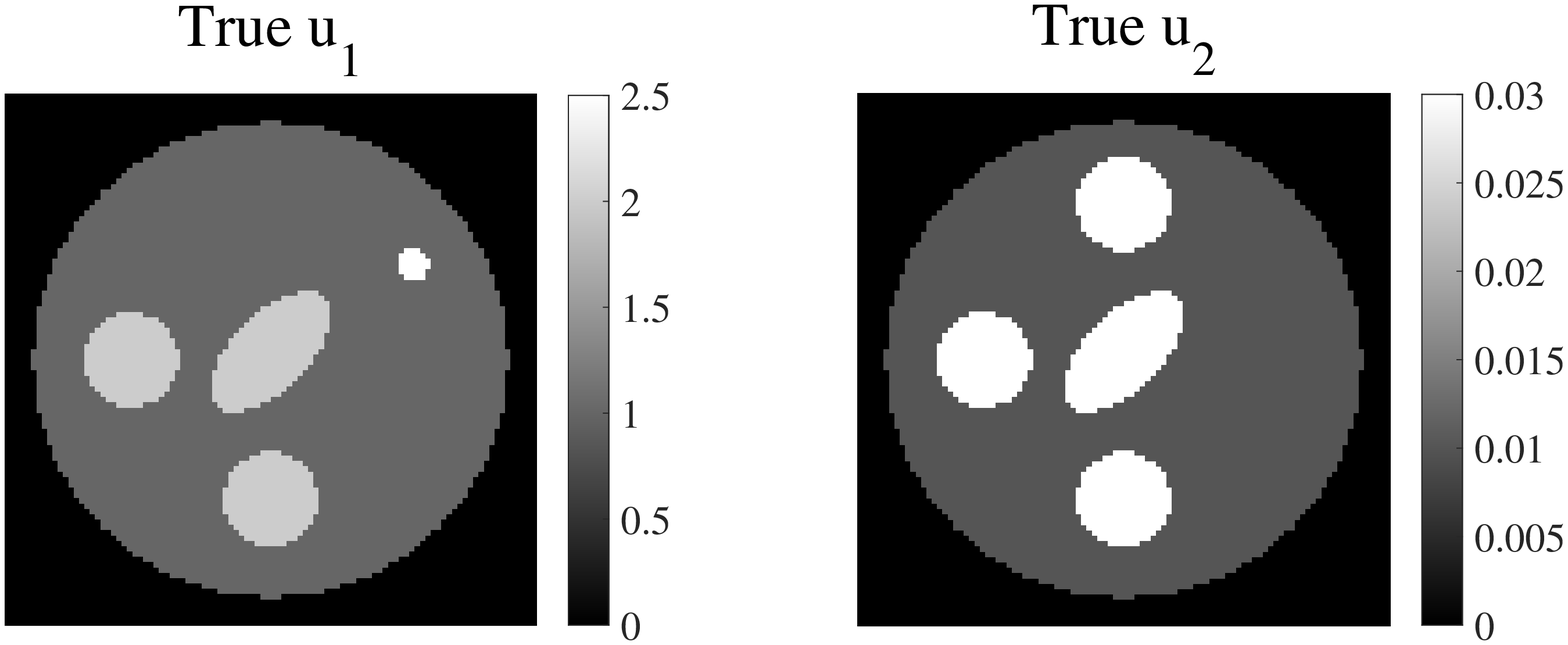}
	\caption{Phantom 3. Left: the true distribution of the X-ray attenuation coefficient $u_1$. Right: the true distribution of the absorption coefficient $u_2$. }
	\label{fig:truedistribution_shape}
\end{figure}

We design three pairs of phantoms of size $100\times 100$. They are all circles with a radius of $25~\text{mm}$. Inside the circles, the structures are shown in Figure \ref{fig:truedistribution}, Figure \ref{fig:truedistribution_same} and Figure \ref{fig:truedistribution_shape}. The X-ray attenuation coefficients $u_1$ and the absorption coefficients $u_2$ are listed in Table \ref{tb:phantom1}, Table \ref{tb:phantom2} and Table \ref{tb:phantom3}.
\begin{table}[htbp]
	\centering
	\caption{Phantom 1}\label{tb:phantom1}
	\begin{tabular}{ccccc}
		\toprule
		& Background& Three circles&Top circle&small circle\\
		\midrule
		XCT~$u_1$~($mm^{-1}$)& 1& 2&---&2.5\\
		DOT~$u_2$~($mm^{-1}$)& 0.01& 0.03&0.03&---\\
		\bottomrule
	\end{tabular}
\end{table}
\begin{table}[htbp]
	\centering
	\caption{Phantom 2}\label{tb:phantom2}
	\begin{tabular}{cccc}
		\toprule
		& Background& Four circles& small circle\\
		\midrule
		XCT~$u_1$~($mm^{-1}$)& 1& 2&2.5\\
		DOT~$u_2$~($mm^{-1}$)& 0.01&0.03&---\\
		\bottomrule
	\end{tabular}
\end{table}
\begin{table}[htbp]
	\centering
	\caption{Phantom 3}\label{tb:phantom3}
	\begin{tabular}{ccccc}
		\toprule
		& Background& Two circle and one ellipse&Top circle&small circle\\
		\midrule
		XCT~$u_1$~($mm^{-1}$)& 1& 2&---&2.5\\
		DOT~$u_2$~($mm^{-1}$)& 0.01& 0.03&0.03&---\\
		\bottomrule
	\end{tabular}
\end{table}
\subsection{Comparison with the SmIR}
To compare the results, we put the reconstructed images of JbmIR-MS and SmIR-MS together with the true distribution, and we adjust the display window all the same as the ground truth, that is $[0,2.5]$ for XCT images and $[0,0.03]$ for DOT images. Since Mumford-Shah functional provides edge information, we also compare the reconstructed edge information using JbmIR-MS and SmIR-MS.
\subsubsection{Example 1}
\begin{figure}[!htbp]
	\centering
	\includegraphics[width = 15cm]{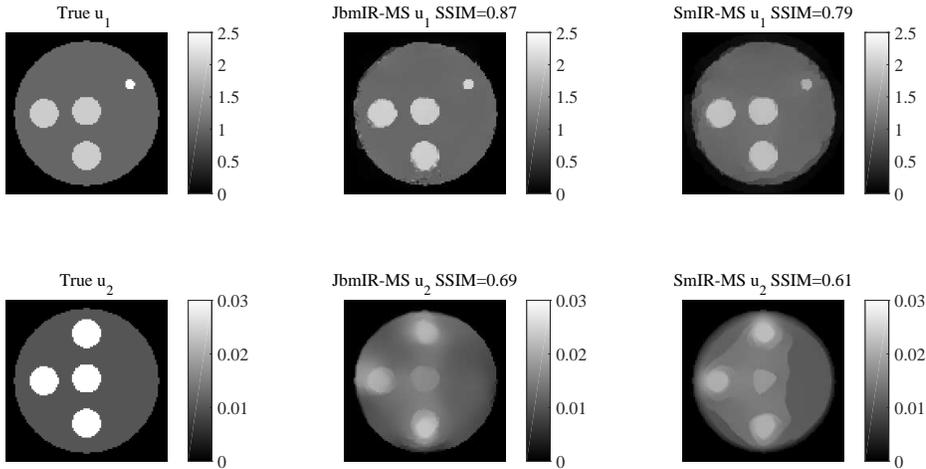}
	\caption{Phantom 1 reconstruction results. (top) The XCT phantom and reconstruction results, (bottom) The DOT phantom and reconstruction results. For SmIR-MS $u_1$ we have SSIM$=0.79$, and for the JbmIR-MS $u_1$ we have SSIM$=0.87$, For SmIR-MS $u_2$, we have SSIM$=0.61$, for JbmIR-MS $u_2$ we have SSIM$=0.69$.  }
	\label{fig:result}
\end{figure}

For JbmIR-MS we choose $\alpha_1=8.8\times10^3,~\beta_1=1.9\times10^{-3},~\gamma_1=9.8,~\alpha_2=1\times 10^5,~\beta_2=6\times 10^{-5},~\gamma_2=5$. For SmIR-MS XCT, we choose $\alpha=8.8\times10^3,~\beta=8\times10^{-3}.$ For SmIR-MS DOT, we choose $\alpha=1\times 10^5,~\beta=5\times 10^{-7}.$ Figure \ref{fig:result} shows the reconstruction results. The three images in the first row from left to right are the true distribution of the X-ray attenuation coefficient $u_1$, JbmIR-MS $u_1$ and SmIR-MS $u_1$. The second row are the true distribution of the absorption coefficient $u_2$, JbmIR-MS $u_2$, and SmIR-MS $u_2$. According to the SSIM values, the image quality of reconstructed $u_1$ is improved using JbmIR-MS (SSIM from 0.79 to 0.87). The image quality of reconstructed $u_2$ is improved using JbmIR-MS (SSIM$=0.69$) then SmIR-MS (SSIM$=0.61$). We notice in the JbmIR-MS results that both modalities maintain their own distinct structure (small circle in XCT and top circle in DOT) and no false structure has been reconstructed because of the information communication with the other modality.

\begin{figure}[!htbp]
	\centering
	\includegraphics[width = 10cm]{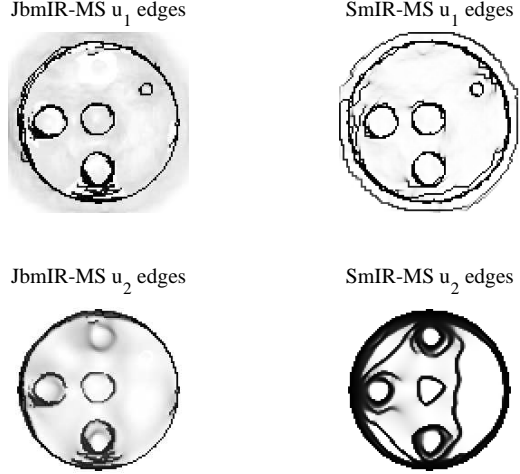}
	\caption{Reconstructed edge images.}
	\label{fig:edge}
\end{figure}

We also show the reconstructed edge images in Figure \ref{fig:edge}. The two images in the first row are reconstructed edge images for X-ray attenuation $u_1$ from bi-modality and single-modality. The second row are reconstructed edge images for absorption coefficient $u_2$ from JbmIR-MS and SmIR-MS. We can find that JbmIR-MS provides clearer edge information than SmIR-MS especially for DOT result. And for JbmIR-MS DOT reconstructed edges, from the difference of the top circle and the other three circles we find that the edge is sharper where XCT image shares the same structure.
\subsubsection{Example 2}
\begin{figure}[!htbp]
	\centering
	\includegraphics[width = 15cm]{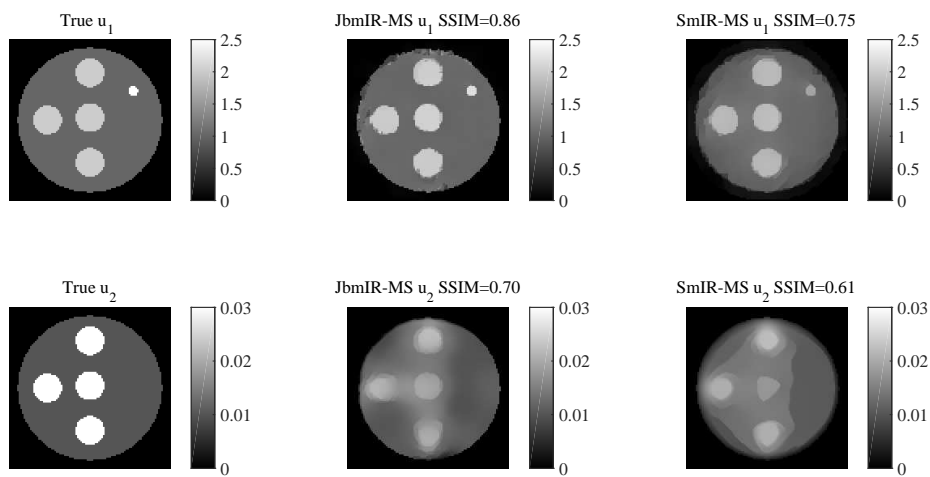}
	\caption{Phantom 2 reconstruction results. (top) The XCT phantom and reconstruction results, (bottom) The DOT phantom and reconstruction results. For SmIR-MS $u_1$ we have SSIM$=0.75$, and for JbmIR-MS $u_1$ we have SSIM$=0.86$, For SmIR-MS $u_2$ we have SSIM$=0.61$, for JbmIR-MS $u_2$ we have SSIM$=0.70$. }
	\label{fig:result_same}
\end{figure}

For JbmIR-MS we choose $\alpha_1=8.8\times10^3,~\beta_1=1.9\times10^{-3},~\gamma_1=9.8,~\alpha_2=1\times 10^5,~\beta_2=7\times 10^{-5},~\gamma_2=5$. For SmIR-MS XCT, we set $\alpha=8.8\times10^3,~\beta=8\times10^{-3}.$ For SmIR-MS DOT, we set $\alpha=1\times 10^5,~\beta=5\times 10^{-7}.$ The reconstructed results are shown in Figure \ref{fig:result_same}. The image quality of reconstructed $u_1$ is improved using JbmIR-MS (SSIM from 0.75 to 0.86). The image quality of reconstructed $u_2$ is higher using JbmIR-MS (SSIM$=0.70$) then SmIR-MS (SSIM$=0.61$). The improvement in DOT image quality is greater than the first example because there are more similar parts between phantoms of two modalities in this example.

\begin{figure}[!htbp]
	\centering
	\includegraphics[width = 10cm]{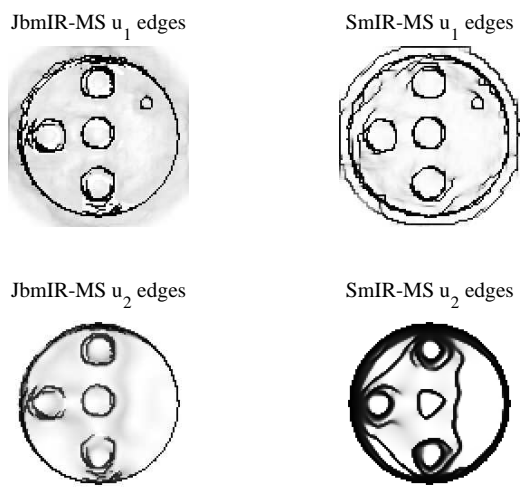}
	\caption{Reconstructed edge images}
	\label{fig:edge_same}
\end{figure}

The edge information is shown in Figure \ref{fig:edge_same}. We can see that the edge information obtained from JbmIR-MS is clearer then SmIR-MS. Compare to the edge results in Example 1, the edge of the top circle for DOT image is much more distinct because the XCT phantom in this example can provide the edge information of the top circle.

\subsubsection{Example 3}
\begin{figure}[!htbp]
	\centering
	\includegraphics[width = 15cm]{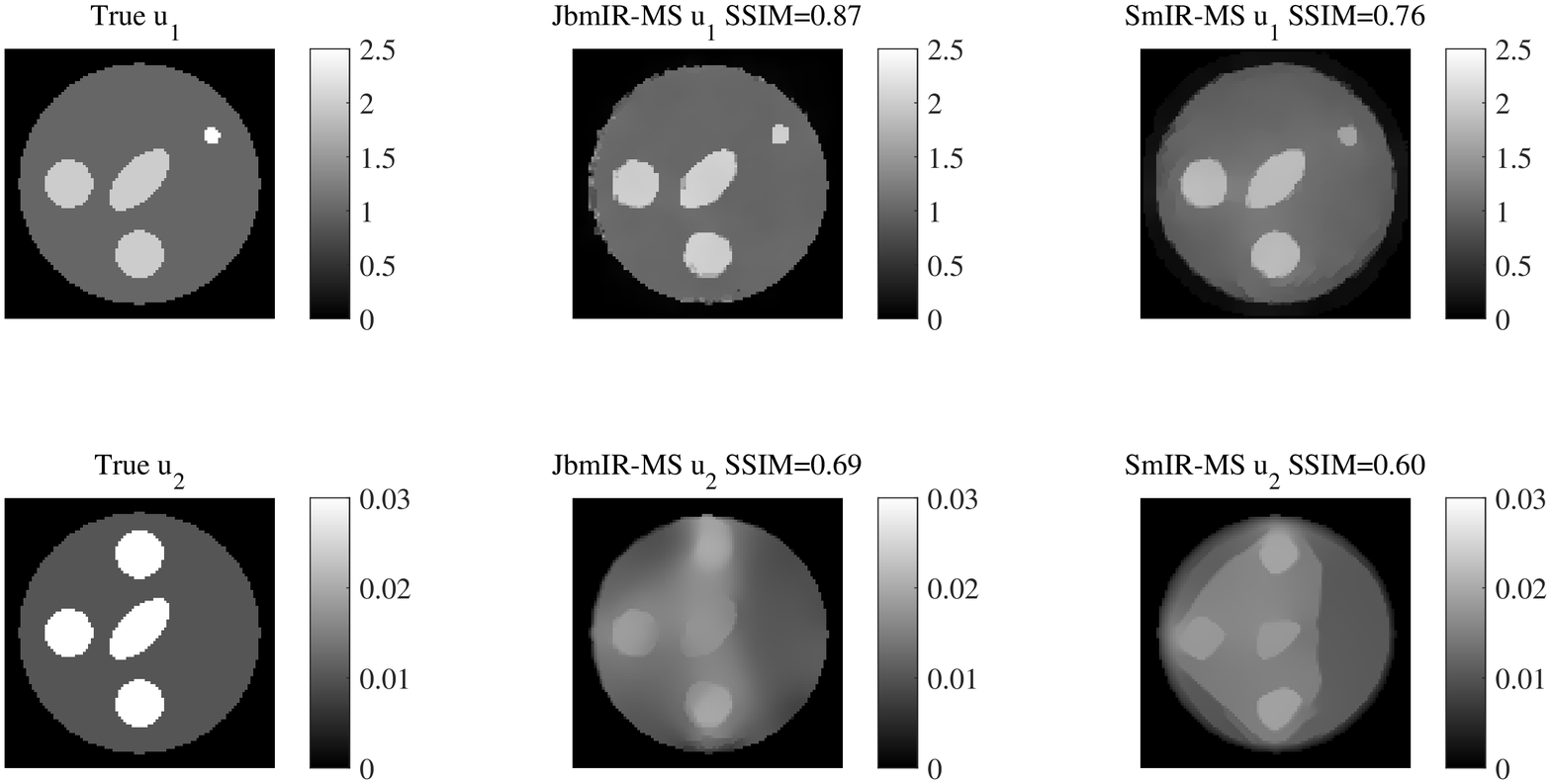}
	\caption{Phantom 3 reconstruction results. (top) The XCT phantom and reconstruction results, (bottom) The DOT phantom and reconstruction results. For SmIR-MS $u_1$ we have SSIM$=0.76$, and for JbmIR-MS $u_1$ we have SSIM$=0.87$, For SmIR-MS $u_2$ we have SSIM$=0.60$, for JbmIR-MS $u_2$ we have SSIM$=0.69$.}
	\label{fig:result_shape}
\end{figure}

For JbmIR-MS we choose $\alpha_1=8.8\times10^3,~\beta_1=1.9\times10^{-3},~\gamma_1=9.8,~\alpha_2=1\times 10^5,~\beta_2=6\times 10^{-5},~\gamma_2=5$. For SmIR-MS XCT, we choose $\alpha=8.8\times10^3,~\beta=8\times10^{-3}.$ For SmIR-MS DOT, we choose $\alpha=1\times 10^5,~\beta=5\times 10^{-7}.$ The reconstruction results are shown in Figure \ref{fig:result_shape}. According to the SSIM values, the image quality of reconstructed $u_1$ is improved using JbmIR-MS (SSIM from 0.76 to 0.87). The image quality of reconstructed $u_2$ is higher using JbmIR-MS (SSIM$=0.69$) then SmIR-MS (SSIM$=0.60$).

\begin{figure}[!htbp]
	\centering
	\includegraphics[width = 10cm]{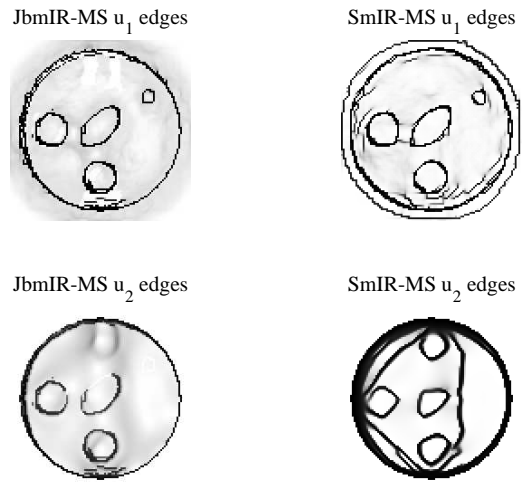}
	\caption{Reconstructed edge images}
	\label{fig:edge_shape}
\end{figure}

The edge information is illustrated in Figure \ref{fig:edge_shape}. We can find that the middle ellipse shape for DOT is more obvious in JbmIR-MS edge result than in SmIR-MS edge result.

\section{Discussion}
\label{sec:discussion}

In this work, we establish an image similarity measure in terms of image edges from Tversky's theory of feature similarity in psychology to solve the DOT and XCT joint reconstruction problem. This similarity consists of Hausdorff measures of the common and different parts of image edges from both modalities. When applied to JbmIR it will improve the reconstructed image quality and will not force the nonexistent structures to be reconstructed.

To implement the joint reconstruction, a $\Gamma-$approximation\eqref{similarity:Psi:Gamma:approx} is proposed for the joint reconstruction functional. The current formulation of the $\Gamma$-approximation is based on the heuristics in \cite[\S~4.5]{Page_2015}. We are unable to prove the expected convergence mathematically, although a proof for $\gamma_1 = \gamma_2 = 0$ might be possible by using the same method in \cite{Page_2015} where a lengthy proof for single modal image reconstruction with the extended Mumford-Shah function with {\it a priori} edge information is provided for $\gamma = 0$. However, it is too tedious to reproduce such a proof in the current work and does not contribute any mathematical insight into this problem with such a stringent condition in our opinion. Instead, we concentrate in studying numerically how the proposed method can help improve the quality of reconstructed images in this work. In the $\Gamma$-approximation functional there isn't any weighting parameters in front of the two fidelity terms, that is because they are independent of each other. We reconstruct the two modalities alternatively, and the information communication is achieved by edge similarity, so we just choose proper regularization parameters for penalty term to control the effect of the other modality and do not weight the fidelity term.

Since circle is the simplest shape, we first design phantoms with circle internal structure. For each example, the structure of the two phantoms are designed similar to each other. XCT should be able to reconstruct finer structure so we add a small circle to the XCT phantoms. DOT cannot distinguish such fine structure so we only design big circle inner structure. Considering that DOT can distinguish different soft tissue better than XCT, we add one more big circle in DOT phantom for Example 1 and Example 3. A reasonable similarity measure should maintain the distinctive \zmark{structure} in each modality and not force the nonexistent features in one modality to be reconstructed. In Example 2, the main parts are set exactly the same as each other to see if we can improve image quality more when the phantoms from two modalities share more common structure. And in Example 3, we change the shape of the middle structure to be an ellipse to see the performance of different methods for different shape of the inner structure.

Our JbmIR-MS is compared with SmIR-MS. From results shown in Figure \ref{fig:result} for Example 1, we can see that the quality of images reconstructed jointly is higher than that of images reconstructed under the SmIR-MS. The SSIM values increase $10.30\%$ and $12.78\%$ for DOT and XCT respectively. Looking at the reconstructed images, JbmIR-MS $u_1$ keeps the structure of the small circle and no extra top circle structure is reconstructed because of the information provided by $u_2$. As for JbmIR-MS $u_2$ the structure is more clear than SmIR-MS result especially the parts that are consistent with $u_1$. It is more obvious in the reconstructed edge information shown in Figure \ref{fig:edge}.

The results for Example 2 \zmark{are} shown in Figure \ref{fig:result_same}. We can see that the image quality improvement is greater than that in Example 1. The SSIM values increase $14.85\%$ and $14.44\%$ for DOT and XCT respectively. We think that the reason that the enhancement in Example 2 is greater than that in Example 1 is because the phantoms of DOT and XCT modalities in Example 2 share more similar parts than Example 1, thus using our edge similarity measure, each modality can provide more \emph{a priori} structural information. Compare to the edge results in Example 1, the edge of the top circle for DOT image is much sharper because the XCT phantom in this example can provide the edge information of the top circle.

In Example 3 we can see that for different shapes of the internal structure, ellipse in our case, JbmIR-MS method also provides better reconstructed images, shown in Figure \ref{fig:result_shape}. The SSIM values increase $13.82\%$ and $15.91\%$ for DOT and XCT respectively. And from the result figure we can clearly see the shape of the middle ellipse in our joint reconstruction result for DOT problem.

\begin{figure}[!htbp]
	\centering
	\subfigure[XCT profiles for Phantom 1.]{
		\label{fig:XCTProfile1}
		\includegraphics[height=6cm]{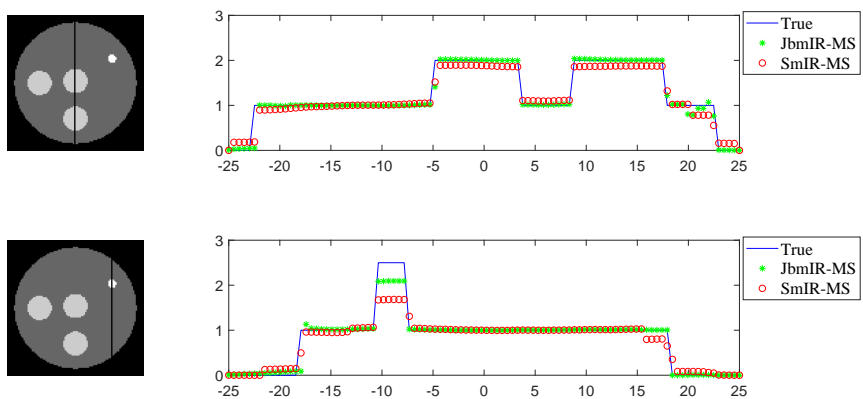}}
	\subfigure[XCT profiles for Phantom 2.]{
		\label{fig:XCTProfile2}
		\includegraphics[height=6cm]{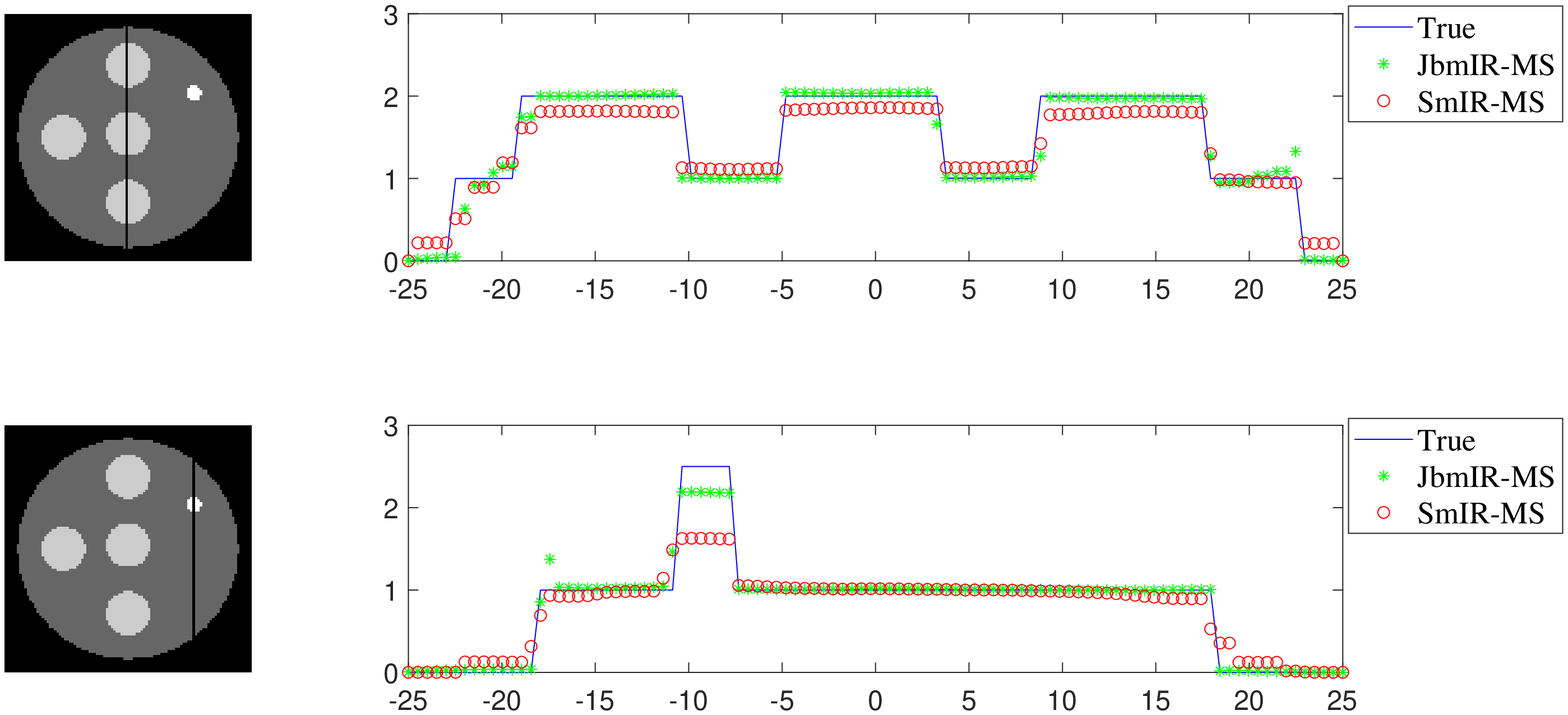}}
	\subfigure[XCT profiles for Phantom 3.]{
		\label{fig:XCTProfile3}
		\includegraphics[height=6cm]{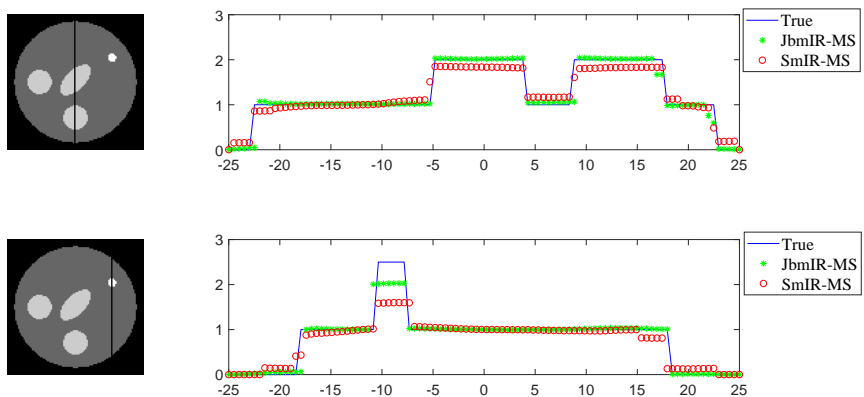}}
	\caption{XCT profiles for the three phantoms and reconstructed images. Black vertical lines on images on the left-side show the positions of lines to draw the line profiles on the right-side. For the line profiles, the blue line shows absorption values of the truth image, red circles shows absorption values from the SmIR-MS of XCT, and green stars shows absorption values from the JbmIR-MS of XCT. It can be seen that the results from the JbmIR-MS \zmark{are} closer to the truth \zmark{distributions} than the SmIR-MS.
	}
	\label{fig:subfig} 
\end{figure}

We find that the visual improvement of XCT reconstruction is not impressive, because the phantoms are easy for XCT. However, looking at their line profiles, the improvement is easy to capture as shown in Figures \ref{fig:XCTProfile1}, \ref{fig:XCTProfile2} and \ref{fig:XCTProfile3}. It can be seen that the XCT results using JbmIR-MS \zmark{are} closer to the true \zmark{distributions} than that using SmIR-MS.

From the numerical results we can see that minimizing the Mumford-Shah functional \zmark{enables} us to reconstruct images and obtain edges at the same time. Since the images from two modalities share similar structure, our JbmIR-MS method adding the edge similarity measure to the Mumford-Shah functional implements the information communication between DOT and XCT modalities which helps to enhance the reconstructed image quality for both modalities. With proper choice of regularization parameters, our similarity measure is able to help \zmark{enhancing} the image quality of the common features, to remain distinctive features for each modality and not force to reconstruct the nonexistent features. The smoothing term together with the usage of separate edge \zmark{variables} to represent structure information enable us to get piecewise smooth reconstructed images with clear and sharp edge with noise data. However there are a number of parameters, including regularization parameters, needed to be chosen carefully and the current parameter choosing is a trial-error approach. For 3D situation, the DOT/XCT bi-modal imaging system has already been set up. We are now preparing the optical phantoms \cite{HeDi2018} for real physical experiment. And we are going to test the JbmIR-MS method with realistic data acquired under our bi-modal imaging system.

\section{Conclusion}
\label{sec:conclusion}

In this paper, we have proposed an asymmetric edge similarity measure following the feature contrast model in \cite{Tversky_1977} to solve the DOT and XCT joint reconstruction problem. This edge similarity measure is able to implement the information communication between the two modalities, thus improves the reconstructed image quality for both modalities. Meanwhile, it remains distinctive structure for each modality and does not force the nonexistent edges to be reconstructed when some of the structures only exist in one modality.

Representing the edge sets by edge indicator functions, a $\Gamma$-approximation \eqref{similarity:Psi:Gamma:approx} is proposed to make it easier to implement the joint reconstruction. To demonstrate the performance of our JbmIR-MS algorithm, we have designed two dimensional numerical phantoms and calculate the minimizers of the $\Gamma$-approximation. To calculate the minimizers, we use gradient \zmark{descent} method with the negative gradient as the \zmark{descent} direction and the step size chosen by backtracking line search with the Armijo-Goldstein stopping condition.

We have evaluated the reconstructed images using SSIM value with the corresponding ground truth as reference image. We compared our JbmIR-MS method with SmIR-MS which shows that joint reconstruction can provide better quality images and clearer edge information.

The results in this work show the effectiveness of the reconstruction method using Mumford-Shah functional with proposed edge similarity measure in 2D.
\section*{Acknowledgments}
This work was partially supported by the Sino-German Center (GZ 1025), National Science Foundation of China (61520106004, 61421062), and National Basic Research Program of China (973 Program) (2015CB351803).

\section*{References}
\bibliographystyle{apalike}
\bibliography{bibitex}
\end{document}